\documentclass[12pt]{article}
\usepackage[dvips]{graphicx}
\usepackage{pdproc}

  \makeatletter 
  \def\@cite#1{[#1]} 
  \newcommand{\ba}{\begin{eqnarray}}
  \newcommand{\ea}{\end{eqnarray}}
  \makeatother    
\begin{document}

\renewcommand{\thefootnote}{\alph{footnote}}

\begin{flushright} \vspace{-2cm} 
{\small MPP-2004-96 \\  
hep-th/0408147} 
\end{flushright}
\vskip 1cm
\title{
 Chiral Gepner Model Orientifolds }

\author{ R. Blumenhagen, T. Weigand }

\address{ Max-Planck-Institut f\"ur Physik \\ 
                  F\"ohringer Ring 6\\
                  D-80805 M\"unchen, Germany 
\\ {\rm E-mail: blumenha@mppmu.mpg.de, weigand@mppmu.mpg.de  }}

\abstract{
We summarize recent progress in constructing orientifolds
      of Gepner models, a phenomenologically interesting class of exactly
      solvable 
     string compactifications with viable gauge groups and chiral matter.}

\normalsize\baselineskip=15pt

\section{Introduction}

The advent of D-branes has not only led to a comprehensive picture of
non-perturbative aspects of string theory and its various dualities, but
has also broadened our view on how our low-energy
world  might be embedded into string theory. In particular, the world-volume
theory on D-branes canonically is a gauge theory so that string
compactifications to four dimensions with D-branes in the background
are interesting models  for string phenomenology.

It is quite intriguing that D-branes naturally not only give rise to gauge
fields but in addition provide a mechanism to yield chiral fermions.
Namely, two non-trivially intersecting D-branes support a chiral
fermion on each intersection point \cite{Berkooz:1996km}. 
Various so-called intersecting D-brane models 
\cite{Blumenhagen:2000wh,Angelantonj:2000hi,Aldazabal:2000cn,Aldazabal:2000dg,Cvetic:2001nr}
have been studied during
the last couple of years, where  mostly the background geometry  was
chosen to be either a torus or a toroidal orbifold 
(see \cite{Uranga:2003pz,Lust:2004ks} for reviews). 
The most promising class of intersecting D-brane models are intersecting
D6-branes on orientifolds of Type IIA string theory. In this case
both the orientifold planes and the D-branes wrap various non-trivially
intersecting 3-cycles of the underlying geometry and fill out the four-dimensional
Minkowskian space-time.

During the last year, techniques have been developed to move beyond this
limited class of toroidal orbifolds and 
to treat also intersecting D-brane models on more general
Calabi-Yau manifolds. Namely, such models were studied 
at very particular symmetric  points in the Calabi-Yau moduli space, where
the non-linear sigma models are  exactly solvable in terms of
superconformal field theories known as Gepner models \cite{Gepner:1987qi}. 
The internal string sector is represented by a tensor product of $N=(2,2)$ worldsheet supersymmetric minimal models. The partition function of each of these 168 theories can be modified by simple current extensions, which yields a huge class of $N=2$ spacetime SUSY string compactifications. The SUSY is broken down to $N=1$ by introducing D-branes and orientifold planes, which are described in terms of boundary and crosscap states respectively. One advantage of this approach is that the CFT is exact to all orders in $\alpha'$. This is particularly important since
geometrically, these models are at radii of the order of the string scale, but without relying on the intuition from classical geometry.   
It would be interesting to study such models
also in the large radius regime \cite{Blumenhagen:2002wn}, where, combined with non-trivial
background fluxes, viable standard like models
with frozen closed string moduli might be possible to achieve. 

Historically, the first fully fledged Gepner model orientifolds
were constructed  in a case by case study  during the nineties 
both in six \cite{Angelantonj:1996mw}  
and four \cite{Blumenhagen:1998tj} uncompactified  dimensions. 
However, during the last year these models saw a revival of interest and
were studied systematically 
\cite{Aldazabal:2003ub,Blumenhagen:2003su,Brunner:2004zd,Blumenhagen:2004cg,Dijkstra:2004ym,Blumenhagen:2004qu,Aldazabal:2004by} (for earlier work on crosscap states in Gepner models
please consult the reference list in these papers).

In this article we briefly summarize the main ingredients for the 
construction of Gepner model orientifolds. Due to lack of space we can explain
neither any technical details nor the notation used in the formulae. 
Therefore, we have to refer the interested reader to
the existing original literature. 

\section{The 1-loop amplitudes}

The massless spectrum can be determined from the various one-loop open string amplitudes, which generically contain divergent tadpole contributions. 
The main result
of the papers \cite{Blumenhagen:2003su,Blumenhagen:2004cg,Blumenhagen:2004qu}
is the computation of the general form of the Klein-bottle and M\"obius amplitude
for A-type simple current extended Gepner model orientifolds,
where the world-sheet parity transformation $\Omega$ may be
dressed by additional phases, 
${\Delta}_j$, and quantum symmetries, $\omega, \omega_\alpha$, \cite{Brunner:2004zd}. Note that via the Greene/Plesser mirror
construction successive orbifolding of the A-type models yields the B-type models, which
have been studied in \cite{Blumenhagen:2003su}. 

In tree channel the Klein-bottle amplitude was shown to be expressible as the overlap
\ba
 \widetilde{K}^A=\int_0^\infty  {dl} \langle C|e^{-2\pi l H_{cl}} |
                C\rangle_A \nonumber 
\ea
with the crosscap state given in the compact form
\ba
\big| C; {\Delta}_j,\omega, \omega_\alpha \big>_{NS} = \frac{1}{\kappa^A_c}
{\sum_{\lambda',\mu'}}^{ev}\, 
\sum_{\nu_0 = 0}^{{K\over 2}-1} \,
\sum_{\nu_j=0}^{1} \,
\sum_{\epsilon_j=0}^{1}\  
(-1)^{\nu_0} \left( \prod_{k<l} (-1)^{\nu_k \nu_l}\right) 
(-1)^{\sum_j \nu_j}\,  (-1)^{\omega \frac{s'_0}{2}}\, \nonumber \\
e^{i \pi \sum_j \frac{\Delta_j}{k_j +2} (m_j' + (1-\epsilon_j)(k_j
 +2))}  \, \left( \prod_\alpha \delta^{(2)}_{Q^{(\alpha)}_{\lambda',
\mu' + (1- \vec{\epsilon})(\vec{k} +2)},\omega_\alpha } \right) 
  \delta^{(4)}_{s'_0 + 2 \nu_0 + 2 \sum \nu_j +2, 2 \omega }\,
\delta^{(2)}_{ \sum_j \frac{1}{k_j +2} (m'_j  + (1-
\epsilon_j)(k_j + 2)), \omega } \nonumber \\ 
\prod_{j=1}^r  \Biggl( \sigma(l_j', m_j', s_j')  
\frac{P_{l'_j, \epsilon_j \, k_j}}{\sqrt{S_{l'_j,0}}} 
 (-1)^{\epsilon_j \frac{m_j' + s_j'}{2}}\, \delta^{(2)}_{m'_j + (1- \epsilon_j)(k_j + 2),0 } \,
 \delta^{(4)}_{s'_j+ 2\nu_0 +2\nu_j + 2(1-\epsilon_j),0} \Biggr)\,\,
\big|{\lambda'},{\mu'}\big>\big>_c. \nonumber 
\ea
As expected the crosscap state is a sum over the crosscap Ishibashi states
weighted with essentially the modular $P=T^{1\over 2}ST^2 ST^{1\over 2}$
matrix of the tensor product of the $N=2$ minimal models subject  to 
extra projections and non-trivial sign factors, which are derived from consistency of the  modular transformation of the M\"obius amplitude. 

In order to cancel the tadpoles from the KB amplitude, we
introduce simple current invariant Cardy-like
A-type boundary states a la Recknagel/Schomerus \cite{Recknagel:1997sb}
\ba
 \big|a\big>_A &=& \big|S_0; (L_j, M_j, S_j)_{j=1}^{r}\big>_A =
\frac{1}{\kappa_{a}^A} 
{\sum_{\lambda',\mu'}}^{\beta}\, 
\prod_{\alpha} \delta^{(1)} \big( Q^{(\alpha)}_{\lambda', \mu'} \big) \nonumber \\
&& (-1)^{\frac{s'^2_0}{2}} e^{ -i\pi \frac{s'_0 S_0}{2}} 
\prod_{j=1}^r \bigg( \frac{S_{l'_j, L_j}}{\sqrt{S_{l'_j,0}}} \,\,
e^{i \pi \frac{m'_j M_j}{k_j +2} } \, e^{-i \pi \frac{s'_j S_j}{2} }\bigg) 
\big|{\lambda',\mu'}\big>\big> . \nonumber 
\ea
(For simple currents with fixed points these boundary states can be refined
into resolved or fractional boundary states.) 

The action of $\Omega_{\Delta_j,\omega,\omega_\alpha}$ on these boundary states
reads
\ba
 |S_0;(L_j,M_j,S_j)_{j=1}^r \rangle_{A} \to
            |-S_0+2\omega;(L_j,-M_j+2\Delta_j,-S_j)_{j=1}^r \rangle_{A} . \nonumber 
\ea
Invariant boundary states carry  $SO(N)$ or  $SP(2N)$ gauge groups, whereas
non-invariant ones have to be introduced in pairs and thus give rise to $U(N)$ gauge groups.

\noindent
A boundary state is supersymmetric relative  to the crosscap state if
\ba
{S_0-\omega\over 2} -\sum_j {M_j-\Delta_j \over k_j+2} + 
          \sum_j {S_j \over 2} =0\ {\rm mod}\ 2 .\nonumber 
\ea
From the crosscap and the boundary states one can compute the 
M\"obius strip amplitude, which together with the annulus amplitude
allows one to read off the open string spectrum.

The tadpole cancellation condition can be easily determined from the contributions of the massless fields $(\lambda, \mu)$ to the amplitudes and takes the form 
\ba
{\rm Tad}_{D}(\lambda,\mu)-4\, {\rm Tad}_{O}(\lambda,\mu)=0.\nonumber 
\ea 
Here, the contribution from the orientifold plane reads 
\ba
{\rm Tad}_{O}(\lambda,\mu) = (-1)^{(1+ \frac{s_0}{2})(1+\omega)} \sum_{\epsilon
\epsilon_r = 0}^1 \, \, 
e^{i \pi \sum_j \frac{\Delta_j}{k_j +2} (1-\epsilon_j)(k_j +2)} 
\left( \prod_{k<l} (-1)^{\epsilon_k \epsilon_l}\right)
\delta^{(2)}_{\sum \epsilon_j, \omega + \frac{s_0}{2}} \nonumber \\ 
 \left( \prod_{\alpha}
\delta^{(2)}_{Q^{(\alpha)}_{\lambda,
\mu + (1- \vec{\epsilon})(\vec{k} +2)},\omega_\alpha} \right) 
\prod_j \Biggl(  {\rm sin}\left[\frac{1}{2}(l_j, \epsilon_j k_j)\right]\,
\delta^{(2)}_{l_j+(1-\epsilon_j) k_j,0}
\delta^{(2)}_{m_j + (1-\epsilon_j)(k_j+2), 0 } \,\,(-1)^{\epsilon_j
 \frac{m_j}{2} } \Biggr). \nonumber 
\ea
Collecting all terms from the boundary states and their 
$\Omega_{{\Delta}_j,\omega,\omega_\alpha}$
images, their massless tadpoles are given by
\ba
{\rm Tad}_{D}(\lambda,\mu)&=&  
{\textstyle \left(  \prod_{\alpha}
\delta^{(1)}_{Q^{(\alpha)}_{\lambda, \mu}}\right) }\,
 \sum_{a=1}^N 2\, N_{a}\, {\rm cos}{\textstyle \left[\pi
     \sum_j \frac{m_j (M_j^a-\Delta_j)}{k_j+2} \right] } \prod_j {\rm sin}(l_j,L_j^a). \nonumber  
\ea

As was shown in \cite{Brunner:2004zd,Blumenhagen:2004cg}, these equations
allow for solutions with unitary gauge symmetries and chiral fermions. 
This is in contrast to the pure B-type orientifolds, which, though easier
to solve, always give rise to non-chiral matter.

\section{Outlook}

We have reviewed the construction of A-type Gepner model orientifolds. 
As a next step, 
it would be interesting to systematically scan all the possible Gepner models for 
phenomenologically interesting examples. First encouraging results
of such a search were reported in  \cite{Dijkstra:2004ym}. 
Moreover for phenomenological applications more refined data is needed and one 
has to determine  the  general form  of  the various terms in the $N=1$ supersymmetric effective action.
It would also be interesting to determine what happens if one moves
away from the Gepner point and, related to this question, to study
such orientifolds in the supergravity regime, where the recently developed methods
of flux compactifications become applicable.

\bibliographystyle{plain}

\begin{thebibliography}{99}
%
\bibitem{Berkooz:1996km}
M.~Berkooz, M.~R.~Douglas and R.~G.~Leigh,
Nucl.\ Phys.\ B {\bf 480}, 265 (1996)
[arXiv:hep-th/9606139].

\bibitem{Blumenhagen:2000wh}
R.~Blumenhagen, L.~G\"orlich, B.~K\"ors and D.~L\"ust,
JHEP {\bf 0010}, 006 (2000)
[arXiv:hep-th/0007024].

\bibitem{Angelantonj:2000hi}
C.~Angelantonj, I.~Antoniadis, E.~Dudas and A.~Sagnotti,
Phys.\ Lett.\ B {\bf 489}, 223 (2000)
[arXiv:hep-th/0007090].

\bibitem{Aldazabal:2000cn}
G.~Aldazabal, S.~Franco, L.~E.~Ibanez, R.~Rabadan and A.~M.~Uranga,
JHEP {\bf 0102}, 047 (2001)
[arXiv:hep-ph/0011132].

\bibitem{Aldazabal:2000dg}
G.~Aldazabal, S.~Franco, L.~E.~Ibanez, R.~Rabadan and A.~M.~Uranga,
J.\ Math.\ Phys.\  {\bf 42}, 3103 (2001)
[arXiv:hep-th/0011073].


\bibitem{Cvetic:2001nr}
M.~Cvetic, G.~Shiu and A.~M.~Uranga,
Nucl.\ Phys.\ B {\bf 615}, 3 (2001)
[arXiv:hep-th/0107166].

\bibitem{Uranga:2003pz}
A.~M.~Uranga,
Class.\ Quant.\ Grav.\  {\bf 20}, S373 (2003)
[arXiv:hep-th/0301032].

\bibitem{Lust:2004ks}
D.~L\"ust,
Class.\ Quant.\ Grav.\  {\bf 21}, S1399 (2004)
[arXiv:hep-th/0401156].


\bibitem{Blumenhagen:2002wn}
R.~Blumenhagen, V.~Braun, B.~K\"ors and D.~L\"ust,
JHEP {\bf 0207}, 026 (2002)
[arXiv:hep-th/0206038].

\bibitem{Gepner:1987qi}
D.~Gepner,
Nucl.\ Phys.\ B {\bf 296}, 757 (1988).


\bibitem{Angelantonj:1996mw}
C.~Angelantonj, M.~Bianchi, G.~Pradisi, A.~Sagnotti and Y.~S.~Stanev,
Phys.\ Lett.\ B {\bf 387}, 743 (1996)
[arXiv:hep-th/9607229].

\bibitem{Blumenhagen:1998tj}
R.~Blumenhagen and A.~Wisskirchen,
Phys.\ Lett.\ B {\bf 438}, 52 (1998)
[arXiv:hep-th/9806131].

\bibitem{Aldazabal:2003ub}
G.~Aldazabal, E.~C.~Andres, M.~Leston and C.~Nunez,
JHEP {\bf 0309}, 067 (2003)
[arXiv:hep-th/0307183].

\bibitem{Blumenhagen:2003su}
R.~Blumenhagen,
JHEP {\bf 0311}, 055 (2003)
[arXiv:hep-th/0310244].

\bibitem{Brunner:2004zd}
I.~Brunner, K.~Hori, K.~Hosomichi and J.~Walcher,
arXiv:hep-th/0401137.

\bibitem{Blumenhagen:2004cg}
R.~Blumenhagen and T.~Weigand,
JHEP {\bf 0402}, 041 (2004)
[arXiv:hep-th/0401148].


\bibitem{Dijkstra:2004ym}
T.~P.~T.~Dijkstra, L.~R.~Huiszoon and A.~N.~Schellekens,
arXiv:hep-th/0403196.


\bibitem{Blumenhagen:2004qu}
R.~Blumenhagen and T.~Weigand,
Phys.\ Lett.\ B {\bf 591}, 161 (2004)
[arXiv:hep-th/0403299].


\bibitem{Aldazabal:2004by}
G.~Aldazabal, E.~C.~Andres and J.~E.~Juknevich,
JHEP {\bf 0405}, 054 (2004)
[arXiv:hep-th/0403262].

\bibitem{Recknagel:1997sb}
A.~Recknagel and V.~Schomerus,
Nucl.\ Phys.\ B {\bf 531}, 185 (1998)
%
\end{thebibliography}

\end{document}